\documentclass[pra,twocolumn,showpacs,eqsecnum,superscriptaddress]{revtex4-1}

\usepackage{graphicx}
\usepackage{verbatim}
\usepackage{color}
\usepackage{amsmath}
\usepackage{amsfonts}
\usepackage[urlcolor=blue, hyperindex, colorlinks, bookmarks=true,linkcolor=black,citecolor=black]{hyperref}
\usepackage{microtype}
\usepackage{xspace}

\usepackage[normalem]{ulem}
\DeclareMathOperator\arcsinh{arcsinh}

\newcommand{\bra}[1]{\left\langle #1\right|}
\newcommand{\ket}[1]{\left|#1\right\rangle}

\newcommand{\mean}[1]{\left\langle #1 \right\rangle}
\newcommand{\trace}[2][]{{\rm Tr_{#1}}\left(#2\right)}
\newcommand{\ImaginaryPart}{{\rm Im}}
\newcommand{\RealPart}{{\rm Re}}

\newcommand{\rp}{\right)}
\newcommand{\lp}{\left(}
\newcommand{\lcb}{\left\{}
\newcommand{\rcb}{\right\}}
\newcommand{\rsb}{\right]}
\newcommand{\lsb}{\left[}
\newcommand{\lbv}{\left|}
\newcommand{\rbv}{\right|}
\newcommand{\lvb}{\lbv}
\newcommand{\rvb}{\rbv}
\newcommand{\inp}[1]{\lp #1 \rp}
\newcommand{\insb}[1]{\lsb #1 \rsb}

\newcommand{\invb}[1]{\lvb #1 \rvb}

\renewcommand{\inf}{\infty}

\newcommand{\pref}[1]{(\ref{#1})}
\renewcommand{\eqref}[1]{Eq.~\pref{#1}}
\newcommand{\eq}[1]{\pref{#1}}
\newcommand{\hc}{\mathrm{h.c.}}

\newcommand{\varlambda}{\Lambda}
\newcommand{\varchi}{X}
\newcommand{\varS}{\mathbb{S}}
\newcommand{\varK}{\mathbb{K}}

\newcommand{\superop}[1]{{\mathcal #1}}
\newcommand{\sD}{{\superop{D}}}
\newcommand{\sL}{{\superop{L}}}

\newcommand{\sQ}{{\superop{Q}}}
\newcommand{\sI}{{\superop{I}}}
\newcommand{\sR}{{\superop{R}}}

\newcommand{\tr}[1]{\mathbf{#1}}

\newcommand{\tS}{{\tr{S}}}
\newcommand{\tD}{{\tr{D}}}

\newcommand{\tP}{{\tr{P}}}

\renewcommand{\wr}{\omega_r}

\newcommand{\projector}{\Pi}
\newcommand{\proj}[1]{\projector_{#1}}

\newcommand{\ad}{a^\dag}
\newcommand{\ada}{a^\dag a}
\newcommand{\sz}{\sigma_0}

\newcommand{\comm}[2]{\lsb #1,#2\rsb}
\newcommand{\paperi}{Paper~I\xspace}

\begin{document}

% Colours
\newcommand{\red}{\color[rgb]{0.8,0,0}}
\newcommand{\green}{\color[rgb]{0.0,0.6,0.0}}
\newcommand{\dkgrn}{\color[rgb]{0.0,0.4,0.0}} %0.6 0.1 0.3
\newcommand{\blu}{\color[rgb]{0,0,0.6}}
\newcommand{\blue}{\color[rgb]{0,0,0.6}}
\newcommand{\pur}{\color[rgb]{0.8,0,0.8}}
\newcommand{\blk}{\color{black}}

\title{Superconducting qubit as a probe of quantum fluctuations in a nonlinear resonator}
\date{\today}

\author{Maxime~Boissonneault}
 \affiliation{D\'epartement de Physique, Universit\'e de Sherbrooke, Sherbrooke, Qu\'ebec, Canada, J1K 2R1}
 \affiliation{Calcul Qu\'ebec, Universit\'e Laval, Qu\'ebec, Qu\'ebec, Canada, G1V 0A6}
 \email{maxime.boissonneault@usherbrooke.ca}
\author{A.~C.~Doherty}
 \affiliation{Centre for Engineered Quantum Systems, School of Physics, The University of Sydney, Sydney, NSW 2006, Australia}
\author{F.~R.~Ong}
\author{P.~Bertet}
 \author{D.~Vion}
\author{D.~Esteve}
\affiliation{Quantronics group, SPEC, IRAMIS, DSM, CEA-Saclay, 91191 Gif-sur-Yvette, France}
\author{A.~Blais}
 \affiliation{D\'epartement de Physique, Universit\'e de Sherbrooke, Sherbrooke, Qu\'ebec, Canada, J1K 2R1}

\begin{abstract}
In addition to their central role in quantum information processing, qubits have proven to be useful tools in a range of other applications such as  enhanced quantum sensing and as spectrometers of quantum noise. Here we show that a superconducting qubit  strongly coupled to a nonlinear resonator can act as a probe of quantum fluctuations of the intra-resonator field. Building on previous work [M.~Boissoneault {\it et al.} Phys. Rev. A {\bf 85}, 022305 (2012)], we derive an effective master equation for the qubit which takes into account squeezing of the resonator field. We show how sidebands in the qubit excitation spectrum that are predicted by this model can reveal information about squeezing and  quantum heating. The main results of this paper have already been successfully compared to experimental data [F.~R.~Ong {\it et al.} Phys. Rev. Lett. {\bf 110}, 047001 (2013)] and we present here the details of the derivations.
\end{abstract}

\maketitle

%&&&&&&&&&&&&&&&&&&&&&&&&&&&&&&&&&
\section{Introduction} % (fold)
\label{sec:introduction}
Nonlinearity in oscillators was first observed by Huygens who discovered that large oscillations in pendulum clocks introduced inaccuracies because of the resulting change in natural oscillation frequency~\cite{Kovacic2011}. It is however most famously Duffing, in the context of combustion engines, who tackled the problem of nonlinear oscillators in a systematic way~\cite{Duffing1918}. Although they have a long history in physics, nonlinear oscillators still manage to surprise and are the focus of intense research~\cite{Boyd2003,Dykman2012}. This is particularly true in optics where optical nonlinearity can be realized by taking advantage of the change in index of refraction of certain media with light intensity. This nonlinearity can lead to frequency down- and up-conversion, and parametric oscillations and amplification~\cite{Boyd2003}. Nonlinearities produced in this context are however rather weak, and nonlinear phenomena at optical frequencies are therefore revealed mostly under high pumping intensities. 

The situation is quite different with superconducting circuits where very strong nonlinearities at microwave frequencies can be achieved~\cite{Bourassa2012}, revealing nonlinear behaviour even at the single-photon level~\cite{lang:2011a,hoffman:2011a,kirchmair:2013a}. These circuits are based on Josephson junctions embedded in otherwise linear circuit elements to create superconducting qubits and nonlinear microwave resonators. These resonators can take various forms, ranging from LC-circuits where the inductance is replaced by a Josephson junction~\cite{Siddiqi2004}, stripline resonators with an embedded Josephson junction~\cite{Boaknin2007}, and to metamaterial resonators where the central resonator conductor is replaced by an array of Josephson junctions~\cite{Castellanos-Beltran2007}. 

These superconducting nonlinear resonators have proven themselves to be valuable tools the context of circuit quantum electrodynamics (cQED) where one couples microwave resonators to superconducting qubits~\cite{Blais2004,Wallraff2004}. In this context, nonlinear resonators have for example been used has parametric~\cite{Vijay2011,riste:2012a,shankar:2013a} or bifurcation amplifiers~\cite{Siddiqi2006a,Lupascu2007,Mallet2009} in qubit state measurement. There, one is interested in the information contained about the state of the qubit in the field at the output of the resonator. In the present paper we take the converse point of view: we show how the qubit can be used as a probe of quantum fluctuations of the field \emph{inside} the resonator.

The theory presented below was developed in parallel to, and already tested against, the experimental results of Ref.~\cite{ong:2013a}. The goal of the present paper is thus to give the details of the derivation of the model whose main results can be found in Ref.~\cite{ong:2013a}. Moreover, the present work is based on the same experimental setup as studied in Ref.~\cite{Ong2011} and builds on previous calculations presented in Ref.~\cite{Boissonneault2012b} -- which we will refer to as \paperi from now on.

In \paperi, we have developed a model describing the measurement backaction of a driven nonlinear resonator on a qubit strongly coupled to the resonator. This model went beyond many approximations that are standard in the literature. First, we considered a many-level instead of a two-level Hilbert space for the superconducting qubit. Second, we took into account the fact that the ac-Stark shift on the qubit caused by a strong pump on the resonator depends on the detuning of the qubit to the pump, and not the qubit-resonator detuning as is usually assumed~\cite{Blais2004}. Finally, our model went beyond the standard linear response theory for the qubit-state dependance of the resonator state. This model was compared with the experimental results presented in Ref.~\cite{Ong2011} and was found to be  in  quantitative agreement with the measured qubit's ac-Stark shift before and after bifurcation of the resonator. We also found excellent agreement with the nontrivial dependence of the qubit's measurement-induced dephasing on the pump power. From these results, we have concluded that the system is close to the quantum limit for measurement in a parameter range.

The model developed in \paperi however made two main approximations that we relax here: small separation between the two pointer states of the resonator corresponding to the two qubit states, and weak squeezing of the resonator field. We still assume that the pointer state separation and the squeezing are both relatively small, but take into account first order corrections to these approximations. As we will show, relaxing the first approximation leads to combined qubit-resonator transitions, i.e. red and blue sideband transitions. Relaxing the second approximation allows us to take into account squeezing of the resonator field and, as predicted by the theory of quantum heating~\cite{Drummond1980,Marthaler2006,dykman:2011a,andre:2012a}, this leads to an effective temperature of the resonator field. Using sideband spectroscopy, a standard tool in ion-trapping experiments~\cite{diedrich:1989a,turchette:2000a}, we then discuss how the qubit can act as an absolute thermometer of this effective temperature. In practice and as discussed in more details below, the main difference between the experimental results of Ref.~\cite{Ong2011} and those of Ref.~\cite{ong:2013a} is a drive of increased amplitude on the qubit.

The production of squeezed light by nonlinear microwave resonators has of course already been studied before~\cite{Yurke1989,Castellanos-Beltran2008}. While these studies focussed on the light at the output of the resonator, as mentioned above here we are focussing on the light inside the resonator.  Moreover, our work adds to an already quite extensive literature concerning nonlinear resonators -- see for example Refs~\cite{Boyd2003,Dykman2012,Dykman1979,Dykman1980,Drummond1980,Dykman2009,Yurke2006,Laflamme2011,Serban2010}. Here however the usual assumptions, such as very small nonlinearities, small qubit-resonator dispersive coupling, or strictly two-level qubits cannot be made when comparing to the experimental results of Ref.~\cite{ong:2013a}, and they are avoided in this paper.

The paper is organized as follow. In section~\ref{sec:presentation_of_the_system}, we present the system's bare Hamiltonian and master equation, and introduce the notation for the nonlinear resonator and the qubit. In section~\ref{sec:summary_of_previous_results}, we summarize the calculation presented in \paperi and highlight the main approximations and results that were obtained. In section~\ref{sec:squeezing_and_sidebands}, we relax the small distinguishability and small squeezing approximations and obtain first order corrections. In section~\ref{sec:sidebands_results}, we compare our model to experimental results first presented in Ref.~\cite{ong:2013a}. Concluding remarks are made in section~\ref{sec:conclusion} while details of some of the calculations can be found in Appendices~\ref{annsec:squeezon_bogoliubov_transformation} and \ref{sec:adiabatic_elimination_through_the_projector_formalism}.

% section introduction (end)
%&&&&&&&&&&&&&&&&&&&&&&&&&&&&&&&&&

%&&&&&&&&&&&&&&&&&&&&&&&&&&&&&&&&&
\section{Presentation of the system} % (fold)
\begin{figure}[t]
	\centering
	\includegraphics[width=\hsize]{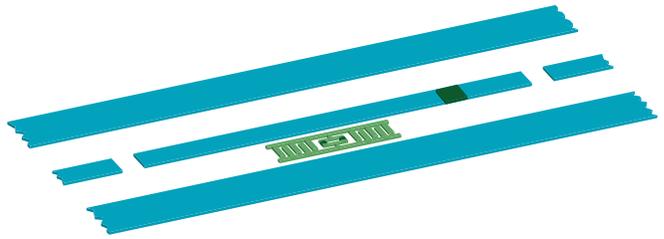}
	\caption{(Color online) Schematic representation of a possible implementation of the system considered in this paper and realized in Ref.~\cite{ong:2013a}. This represents a coplanar resonator (blue) made nonlinear using an embedded Josephson junction (dark green) and capacitively coupled to a transmon qubit~\cite{Koch2007} placed between the central conductor and the ground planes. The model described in this paper however applies to other nonlinear resonators and qubits (see text). }
	\label{fig:circuit_transmon}
\end{figure}

\label{sec:presentation_of_the_system}
As discussed in the Introduction, we consider a nonlinear resonator strongly coupled to a superconducting qubit. An example of such a system is illustrated in Fig.~\ref{fig:circuit_transmon}, where a transmon qubit~\cite{Koch2007} is coupled to a coplanar transmission-line resonator rendered nonlinear by Josephson junction embedded in the resonators' center conductor.  We introduce in section~\ref{sub:nonlinear_resonator} the notation used for the nonlinear resonator and in section~\ref{sub:qubit_and_coupled_system} we focus on the qubit and its coupling to the resonator.

%&&&&&&&&&&&&&&&&&&&&&&&&&&&&&&&&&
\subsection{Nonlinear resonator} % (fold)
\label{sub:nonlinear_resonator}
Following the notation of Yurke and Buks~\cite{Yurke2006} and also used in \paperi, we define the Hamiltonian of the nonlinear resonator as ($\hbar=1$) 
\begin{equation}
	\label{eqn:H_r}
	H_r = \wr\ada + \frac{K}{2} \ad\ad aa + \frac{K'}{3} {\ad}^3 a^3,
\end{equation}
where $\omega_r$, $K$ and $K'$ are respectively the resonator's bare resonance frequency, its Kerr coefficient and a higher order nonlinearity Kerr coefficient. The operator $a^{(\dag)}$ annihilates (creates) an excitation in the resonator.

An important aspect of the experiment described in Ref.~\cite{ong:2013a} is the presence of multiple drives on the resonator. We will denote with the subscript $d$ any drive, of amplitude $\epsilon_d$ and frequency $\omega_d$, far detuned from the qubit transition frequency. We will allow for many such drives in our description. In addition, we will denote with the subscript $s$ a spectroscopy drive of amplitude $\epsilon_s$ and frequency $\omega_s$ close to the qubit's transition frequency. The presence of these drives can be represented by the usual Hamiltonians~\cite{Gardiner2000},
\begin{subequations}
	\begin{align}
		\label{eqn:H_d}
		H_d &= \sum_{d} \left(\epsilon_d e^{-i\omega_d t} \ad + \epsilon_d^* e^{i\omega_d t} a\right), \\
		H_s	&= \epsilon_s e^{-i\omega_s t}\ad + \epsilon_s^* e^{i\omega_s t} a.
	\end{align}
\end{subequations}
The drives $d$, far from the qubit resonance, are used to populate the resonator and will not drive transitions of the qubit. They will result in dispersive shifts of the qubit frequency. On the other hand, the spectroscopy drive is aimed specifically at driving the qubit. Because of their different influence, we treat these various drives very differently below. 

We finally introduce photon loss in the resonator at the rate $\kappa$. Together with the above Hamiltonian, this is captured by the Lindblad form master equation
\begin{equation}
	\label{eqn:master_equation_resonator}
	\dot\rho_r \equiv \sL_r \rho_r = -i\comm{H_r+H_d+H_s}{\rho_r} + \kappa\sD[a]\rho_r,
\end{equation}
where we have introduced the usual dissipation superoperator
\begin{equation}
	\label{eqn:dissipator}
	\sD[A]\rho \equiv \frac{1}{2} (2A\rho A^\dag - A^\dag A\rho - \rho A^\dag A).
\end{equation}

\begin{figure}
	\centering
	\includegraphics[width=\hsize]{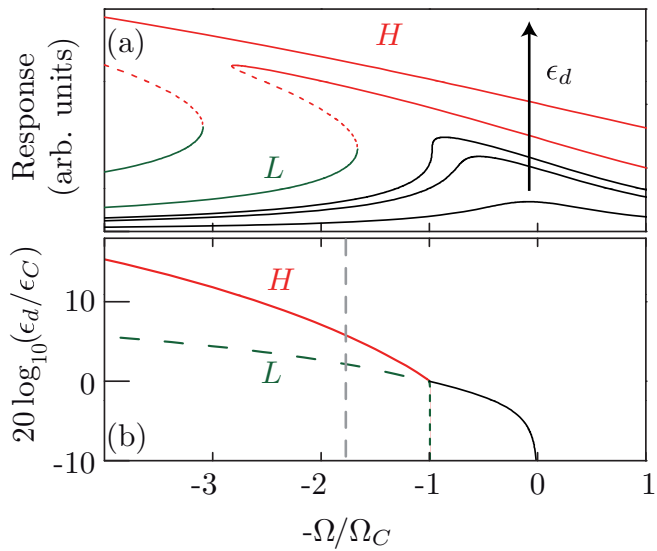}
	\caption{ (Color online) (a) Amplitude of the resonator internal field (arbitrary units) in response to a drive of reduced frequency $\Omega=2(\omega_r-\omega_d)/\kappa$ ($\Omega_C=\sqrt{3}$) for increasing drive amplitudes $\epsilon_d$. (b) Stability diagram of the resonator. In the region $H$ (above the full line), the resonator is in a high-amplitude state. In the region $L$, (below dashed line), it is in a low-amplitude state. Between the dashed and full lines, the resonator is bistable. The dashed vertical line corresponds to the detuning at which the data presented in this paper were taken. }
	\label{fig:bifurcation_diagram}
\end{figure}

Nonlinear resonators in circuit QED typically have a negative Kerr constant $K$~\cite{Bourassa2012}. As one drives the resonator, the nonlinearity therefore causes a back-bending of the resonator's response as illustrated in Fig.~\ref{fig:bifurcation_diagram}~(a). For large drive amplitudes $\epsilon_d$, the resonator becomes bistable, with the two stable solutions denoted $L$ and $H$, and of respectively low and high amplitude of oscillation. To simplify the description, it is useful to introduce the reduced detuning $\Omega = 2(\omega_r-\omega_d)/\kappa$~\cite{Ong2011}. As a function of this reduced detuning and of the drive amplitude $\epsilon_d$, the resonator's response is captured by the stablity diagram illustrated in Fig.~\ref{fig:bifurcation_diagram}~(b). If $\Omega$ is smaller than the critical detuning $\Omega_C = \sqrt{3}$ the resonator can be in the $L$ state at low power and in the $H$ state at high power, with an intermediate region where both states are stable. Since both states go from stable to unstable at different powers, the resonator in this region is hysteretic, and can be used as a sample-and-hold detector~\cite{Siddiqi2006,Lupascu2007,Mallet2009}. 
% subsection nonlinear_resonator (end)
%&&&&&&&&&&&&&&&&&&&&&&&&&&&&&&&&&

%&&&&&&&&&&&&&&&&&&&&&&&&&&&&&&&&&
\subsection{Qubit and qubit-resonator coupling} % (fold)
\label{sub:qubit_and_coupled_system}
Most superconducting qubits require a larger Hilbert space than their logical subspace $\{\ket0,\ket1\}$ for an accurate description. This is the case for example for the transmon~\cite{Koch2007}, the capacitively-shunted flux qubit~\cite{Steffen2010}, the fluxonium~\cite{Manucharyan2009}, the tunable-coupling qubit~\cite{Gambetta2011} and the phase qubits~\cite{Martinis2009}. Here, we consider up to $M$ levels and write the free qubit Hamiltonian as 
\begin{equation}
	\label{eqn:H_q}
	H_q = \sum_{i=0}^{M-1}\omega_i\proj{i,i} \equiv \proj{\omega},
\end{equation}
where $\omega_i$ is the frequency of the qubit eigenstate $\ket{i}$, $\proj{i,j}\equiv\ket{i}\bra{j}$, and where we have introduced the short-handed notation
\begin{equation}
	\label{eqn:proj_x}
	\proj{x} \equiv \sum_{i=0}^{M-1} x_i \proj{i,i}.
\end{equation}
In practice, the values of the eigen-frequencies $\omega_i$ should either be extracted from experiments, or computed by diagonalizing the full qubit Hamiltonian. For the results presented in this paper, we considered a transmon qubit and carefully calibrated the qubit frequencies from experimental spectroscopic data. More details about the sample and its parameters can be found in Refs.~\cite{Ong2011,ong:2013a} and in \paperi.

As usual, we consider a dipolar qubit-resonator coupling
\begin{equation}
	\label{eqn:H_I}
	H_I = \sum_{i=0}^{M-2} g_i (\ad+a) (\proj{i,i+1} + \proj{i+1,i}),
\end{equation}
where each qubit transition $i\leftrightarrow j$ is coupled to the resonator if and only if $i = j \pm 1$. This restrictive condition is often made true either by selection rules or because the other transitions are too far detuned from the resonator frequency to have an impact~\cite{Koch2007}. We stress that we use here the full Rabi Hamiltonian rather than its Jaynes-Cummings counter-part since, as will be seen below, the counter-rotating terms in~\eqref{eqn:H_I} will play a predominant role in the sideband transitions.

Taking into account qubit damping and pure dephasing, we finally write the master equation describing the coupled system as 
\begin{equation}
	\begin{split}
		\label{eqn:master_equation}
		\dot\rho &= -i\comm{H}{\rho} + \kappa\sD[a]\rho \\
		&\quad + \gamma \sum_{i=0}^{M-2} \lp\frac{g_i}{g_0}\rp^2 \sD\lsb \proj{i,i+1} \rsb\rho + 2\gamma_\varphi \sD\lsb \proj{\varepsilon} \rsb\rho,
	\end{split}
\end{equation}
where the total Hamiltonian is $H = H_r+H_q+H_I+H_d+H_s$. In this master equation, $\gamma$ is the qubit $\ket{1}\rightarrow \ket{0}$ decay rate and $\gamma_\varphi$ is the qubit pure dephasing rate for the same states. With the above form, we have assumed that qubit decay between two consecutive states scales as the coupling of this transition to the resonator.  This assumption is not essential to this work but is convenient and realistic. Moreover, to describe pure dephasing of this multi-level system, we have defined $\proj{\varepsilon}$ with $\varepsilon_i \equiv \frac{\partial (\omega_i - \omega_0)}{\partial X} \times \left[ \frac{\partial(\omega_1-\omega_0)}{\partial X}\right]^{-1}$ the $X$-dispersion and with $\varepsilon_0=0$, $\varepsilon_1 = 1$ by definition.. Here, $X$ represents some control parameter (for example flux or charge) whose fluctuations cause dephasing. %As with most master equations, this master equation assumes white noise of the baths spectra and weak coupling to the environment. 
The above master equation is the same as the one used in \paperi, with the exception that we have set here the resonator's rate of two-photon loss~\cite{Yurke2006} to zero for simplicity~\footnote{Since we are disregarding two-photon losses, we have that $\kappa'''$ in the notation of \paperi is here simply equal to $\kappa$.}. 

Our goal in the next two sections is to obtain a reduced qubit model that includes squeezing of the resonator field and captures qubit-resonator sideband transitions in the high-power regime of the spectroscopy drive. In section~\ref{sec:summary_of_previous_results}, we first summarize the results obtained in \paperi. In section~\ref{sec:squeezing_and_sidebands}, we then build on these results and consider first order corrections to two main approximations that were used in \paperi and mentioned in the Introduction. We will show that in the presence of strong spectroscopy drive $\epsilon_s$, these corrections will yield a qubit spectrum displaying red and blue sidebands in addition to the main qubit line, and that the amplitude of these sidebands reveals information about squeezing of the intra-resonator field. 
% subsection qubit_and_coupled_system (end)
%&&&&&&&&&&&&&&&&&&&&&&&&&&&&&&&&&

% section presentation_of_the_system (end)
%&&&&&&&&&&&&&&&&&&&&&&&&&&&&&&&&&

%&&&&&&&&&&&&&&&&&&&&&&&&&&&&&&&&&
\section{Summary of previous results} % (fold)
\label{sec:summary_of_previous_results}
In \paperi, we have performed a series of unitary transformations on the master equation \eq{eqn:master_equation} and have obtained an effective master equation for the qubit only. The first step is to transform the master equation using a polaron transformation~\cite{Gambetta2008}
\begin{equation}
	\label{eqn:polaron_transformation}
	\tP = \sum_{i=0}^{M-1} \proj{i,i} D(\alpha_i),
\end{equation}
where $D(\alpha) = \exp[\alpha \ad - \alpha^* a]$ is the displacement operator~\cite{Gardiner2000}.  This transformation displaces the resonator field in a qubit-state-dependent manner, such that $a \rightarrow a + \proj{\alpha}$. If the pointer states $\alpha_i$ are chosen properly, the intra-resonator field in this transformed frame is in --- or close to --- the vacuum. In this situation, it is simple to trace over the resonator states to obtain an effective equation for the qubit only. This can be done exactly within the linear dispersive approximation~\cite{Gambetta2008}, but unfortunately not when taking into account the full Jaynes-Cummings coupling such as in \paperi. In this situation, the additional complexity arises from transforming the operator $\proj{i,i+1}$ in \eqref{eqn:H_I} which yields
\begin{equation}\label{eq:PolaronOnPi}
	\tP^\dag \proj{i,i+1}\tP = \proj{i,i+1} D(\alpha_{i+1}-\alpha_{i}) e^{-i\ImaginaryPart[\alpha_{i+1}^*\alpha_i]}.
\end{equation}
This transformed operator is problematic since it contains all powers of the ladder operators $a^{(\dag)}$ throught the displacement operator and these will not leave the resonator field in its vacuum state in the transformed frame. To simplify the situation we assumed in \paperi that the distinguishability $|\alpha_{i+1}-\alpha_i|$ is very small and took $\tP^\dag \proj{i,i+1} \tP \approx \proj{i,i+1}$. With this approximation, the interaction Hamiltonian $H_I$ transforms into a detuned drive acting directly on the qubit. 

The second step in \paperi is to remove this effective detuned qubit driving using what we called a classical dispersive transformation
\begin{equation}
	\label{eqn:classical_dispersive_transformation}
	\tD_C = \exp\lsb \sum_{i=0}^{M-2} \xi_i^* \proj{i,i+1} - \xi_i \proj{i+1,i}\rsb,
\end{equation}
where $\xi_i$ is a classical (scalar) analog of the operator $\lambda_i \ad = [g_i/(\omega_{i+1,i}-\omega_r)] \ad$ found in the usual dispersive transformation of the Jaynes-Cummings Hamiltonian~\cite{Boissonneault2009}. 

After these two steps, the result is a transformed master equation containing the ac-Stark shift of the qubit frequency, dressed-dephasing of the qubit~\cite{Boissonneault2008} and measurement-induced dephasing~\cite{Gambetta2006}. These various quantities are related by the nonlinear equations for the pointer states
\begin{equation}
	\label{eqn:condition_alpha_d}
	\begin{split}
		0 &= (\omega_r-\omega_d-i\tfrac{\kappa}{2})\alpha_{i,d} + K|\alpha_i|^2\alpha_{i,d} \\
		&\quad  + K'|\alpha_{i}|^4\alpha_{i,d} + \epsilon_d + \inp{\varS_i^d+\frac{1}{3!}\varK_i^d|\alpha_i|^2} \alpha_{i,d},
	\end{split}
\end{equation}
where the expressions for $\varS_i^d$ and $\varK_i^d$ are given below. With this formulation of \eqref{eqn:condition_alpha_d} we have assumed that $\alpha$ can be written as $\alpha = \sum_i \alpha_i =  \sum_{d,i} \alpha_{i,d} e^{-i\omega_d t} + \alpha_{i,s} e^{-i\omega_s t}$. This form assumes that the multiple drives are spread out enough in frequency such that one drive does not contribute significantly to the field oscillating at another drive's frequency. 

In the third step, we apply one last transformation, the quantum dispersive transformation, which takes here the form
\begin{equation}
	\label{eqn:quantum_dispersive_transformation}
	\tD = \exp\insb{ \sum_{i=0}^{M-2} \lambda_i \ad \proj{i,i+1} - \lambda_i^* a \proj{i+1,i}},
\end{equation}
with $\lambda_i = g_i/(\omega_{i+1,i}-\omega_r)$. Since the polaron transformation moves the system to a frame where the photon population is small, this transformation can safely be performed to lowest order. The resulting master equation now contains the Lamb shift of the qubit frequency as well as Purcell decay. 

After these three transformations and projecting the qubit into its logical subspace $\{\ket{0},\ket{1}\}$, the effective  Hamiltonian takes the form $H''' = H_0'''+H_2'''$ (each prime indicating a transformation) where 
\begin{subequations}
	\begin{align}
		\label{eqn:H_0_third}
		H_0''' &= \frac{\omega_{1,0}'''}{2}\sigma_z + g_0\inp{\alpha_{0,s} e^{-i\omega_s t} \sigma_+ + \hc}, \\
		\label{eqn:H_2_third}
		H_2''' &= \lsb\omega_r'(\alpha) + \proj{S(\alpha)}\rsb\ada + \Upsilon{\ad}^2 + \Upsilon^*a^2.
	\end{align}
\end{subequations}
In these expressions, the ac-Stark and Lamb shifted qubit frequencies are given by
\begin{equation}
	\label{eqn:omega_i_third}
	\begin{split}
		\omega_{i}'''(\alpha) &\equiv \omega_i''(\alpha) + L_i(\alpha), \\
		\omega_{i}''(\alpha) &\equiv \omega_i + \sum_{d} \varS_i^d |\alpha_d|^2 + \frac{1}{4} \sum_{d} \varK_i^d|\alpha_d|^4		
	\end{split}
\end{equation}
where
\begin{equation}
	\label{eqn:classical_stark_shift_coefficients}
	\begin{split}
		\varS_i^d &\equiv -(\varchi_i^d - \varchi_{i-1}^d), \\
		\varK_i^d &\equiv -4\varS_i^d(|\varlambda_i^d|^2+|\varlambda_{i-1}^d|^2) \\
		&\quad - (3\varchi_{i+1}^d|\varlambda_i^d|^2-\varchi_i^d|\varlambda_{i+1}^d|^2) \\
		&\quad + (3\varchi_{i-2}^d|\varlambda_{i-1}^d|^2-\varchi_{i-1}^d|\varlambda_{i-2}^d|^2), 
	\end{split}
\end{equation}
are the classical Stark shift coefficients, with $\varlambda_i^d \equiv -g_i/(\omega_{i+1,i}-\omega_d)$ and $\varchi_i^d \equiv -g_i\varlambda_i^d$, and where
\begin{subequations}
	\begin{align}
		L_i(\alpha) &\equiv \chi_{i-1}(\alpha), \\
		S_i(\alpha) &\equiv -(\chi_i(\alpha) - \chi_{i-1}(\alpha)),
	\end{align}
\end{subequations}
are the Lamb shift and the cavity-pull, with $\chi_i(\alpha) \equiv -g_i\lambda_i(\alpha)$ and $\lambda_i(\alpha) \equiv -g_i/[\omega_{i+1,i}''(\alpha) - \omega_r'(\alpha)]$. We have also defined
\begin{equation}
	\Upsilon \equiv \inp{\frac{K}{2} + K'|\proj{\alpha}|^2} \proj{\alpha}^2.
\end{equation}
Because of its $a^{(\dag)2}$ dependence, the term proportional to $\Upsilon$ in the transformed Hamiltonian leads to squeezing of the resonator field. This contribution was assumed to be small in \paperi and dropped. 

Putting all of this together, the resulting transformed qubit-resonator master equation is then given by
\begin{equation}
	\label{eqn:master_equation_third}
	\begin{split}
		\dot\rho''' &= -i\comm{H_0'''+H_2'''}{\rho'''} + \kappa\sD[a]\rho''' \\
		&\quad + \gamma_\downarrow''' \sD[\sigma_-]\rho''' + \frac{\gamma_\varphi'''}{2} \sD[\sigma_z]\rho''',
	\end{split}
\end{equation}
where
\begin{subequations}
	\begin{align}
		\gamma_\downarrow''' &= \gamma + \lambda_0^2(\alpha)\kappa, \\
		\gamma_\varphi''' &= \gamma_\varphi + \Gamma_{\varphi m}, \\
		\label{eqn:gamma_varphim}
		\Gamma_{\varphi m} &= \frac{\kappa |\alpha_1-\alpha_0|^2}{2},
	\end{align}
\end{subequations}
are the modified rates having neglected dressed dephasing as well as two-photon losses, both of which were included in \paperi.

Using this master equation, we have showed in \paperi that the measurement-induced dephasing rate given by~\eqref{eqn:gamma_varphim}, with the pointer states given by~\eqref{eqn:condition_alpha_d}, is in quantitative agreement with measured qubit spectroscopic linewidth within the limits of the approximations that were made. There are few key points behind this good agreement between theory and experiments. First, contrary to what is usually used in circuit QED, our expression for the ac-Stark shift depends on the pump drive frequency rather than the resonator frequency. Our treatment moreover goes beyond linear response theory for the resonator state. Indeed, in \paperi we show that for typical circuit QED parameters, describing the measurement-induced dephasing quantitatively with a nonlinear resonator always requires going beyond a linear response. That is, whenever the gain of the resonator is large, the susceptibility of the resonator response to a shift in its resonance frequency is large. Because of this large susceptibility, the qubit cannot be treated as a simple perturbation causing a small shift of the resonator frequency.

% section summary_of_previous_results (end)
%&&&&&&&&&&&&&&&&&&&&&&&&&&&&&&&&&

%&&&&&&&&&&&&&&&&&&&&&&&&&&&&&&&&&
\section{Squeezing and sidebands} % (fold)
\label{sec:squeezing_and_sidebands}
In this section, we relax the two main approximations that are discussed above (small distinguishability and negligible squeezing) and include first order corrections. First, instead of approximating \eqref{eq:PolaronOnPi} as $\tP^\dag \proj{i,i+1} \tP \approx \proj{i,i+1}$, we now take
\begin{equation}
	\tP^\dag \proj{i,i+1}\tP \approx \proj{i,i+1} \inp{ 1 + \beta_i \ad - \beta_i^* a},
\end{equation}
where $\beta_i \equiv \alpha_{i+1}-\alpha_i$ is assumed to be small. Taking these terms into account when transforming $H_I$ yields a term in the Hamiltonian that was neglected in \paperi and that is given by
\begin{equation}
	\label{eqn:H_SB}
	H_\mathrm{SB}' \equiv \sum_{i=0}^{M-2} g_i \inp{\proj{\alpha}^*+\proj{\alpha}} \insb{\inp{\beta_i \ad - \beta_i^*a}\proj{i,i+1}+\hc}.
\end{equation}
This Hamiltonian generates multi-photon qubit-resonator sideband transitions that will appear in the qubit spectrum.

With the proper choice of polaron frame (i.e. of pointer states $\alpha_i$), the average field $\mean{a}$ is small. Assuming that $\beta_i$ is also small, we consider $H_\mathrm{SB}'$ to be itself a correction to the transformed system Hamiltonian. We therefore omit to apply the classical and quantum dispersive transformations on $H_\mathrm{SB}'$ since this would only yield even smaller corrections.

Taking into account this correction, the master equation \eq{eqn:master_equation_third} describing the system in the three-times transformed frame now reads
\begin{equation}
	\label{eqn:master_equation_third_sb}
	\begin{split}
		\dot\rho''' &= -i\comm{H_0'''+H_2''' + H_\mathrm{SB}'''}{\rho'''} + \kappa\sD[a]\rho''' \\
		&\quad + \gamma_\downarrow''' \sD[\sigma_-]\rho''' + \frac{\gamma_\varphi'''}{2} \sD[\sigma_z]\rho''',
	\end{split}
\end{equation}
where $H_{\mathrm{SB}}''' = H_{\mathrm{SB}}'$.

%&&&&&&&&&&&&&&&&&&&&&&&&&&&&&&&&&
\subsection{Bogoliubov transformation} % (fold)
\label{sub:bogoliubov_transformation}

 In \paperi, we had only $H_0''' + H_2'''$ as the Hamiltonian in the transformed frame and assumed that $H_2'''$ was a perturbation small enough that the state of the resonator was the vacuum and could be readily traced out. In practice however the terms proportional to $a^{(\dag)2}$ in $H_2'''$ lead to squeezing of the stationary state of the resonator. Here, we do not drop these terms and take care of them using  a Bogoliubov transformation before adiabatically eliminating the resonator.

Indeed, assuming Hamiltonian $H_2'''$ of \eqref{eqn:H_2_third} depends only weakly on the qubit state, it can be diagonalized using a Bogoliubov transformation, which takes the form
\begin{equation}
	\label{eqn:tS}
	\tS = e^{\tfrac12 \xi^* a^2 - \tfrac12 \xi {\ad}^2}
\end{equation}
and whose action on the field operator $a$ is~\cite{Gardiner2000}
\begin{equation}
	\label{eqn:a_S}
	\tS^\dag a \tS = \cosh(r) a - e^{i2\theta}\sinh(r) \ad,
\end{equation}
where $\xi = r e^{2i\theta}$ is the squeezing parameter.

To keep the analytical calculations tractable, we will keep in this transformation only the time-dependance of $\theta$ with $\theta(t) \equiv -\omega_p t + \Theta$. This implies that the field has reached a steady state in a frame rotating at $\omega_p$. We show in Appendix~\ref{annsec:squeezon_bogoliubov_transformation} that under transformation by $\tS$ the qubit-resonator master equation now takes the form
\begin{equation}
	\label{eqn:master_equation_4}
	\begin{split}
		\dot\rho^{(4)} &= -i\comm{H_s^{(4)}}{\rho^{(4)}} + \gamma_{\downarrow}''' \sD\lsb \sigma_- \rsb\rho^{(4)} + \frac{\gamma_\varphi'''}{2} \sD[\sigma_z]\rho^{(4)} \\
		& + \kappa\lsb\sinh^2(r)+1\rsb\sD[a]\rho^{(4)} + \kappa\sinh^2(r)\sD[\ad]\rho^{(4)},
	\end{split}
\end{equation}
where $H_s^{(4)} = H_0^{(4)} + H_2^{(4)} + H_{\mathrm{SB}}^{(4)}$, with
\begin{equation}
	\begin{split}
		H_0^{(4)} &= g_0 \inp{\alpha_{s,0}\sigma_+ e^{i\delta t} + \hc}, \\
		H_2^{(4)} &= \tilde\Delta_r(\alpha) \inp{\ada+\tfrac12}, \\
		H_{\mathrm{SB}}^{(4)} &= F^{(4)} \sigma_+e^{i\delta t} + \hc,
	\end{split}
\end{equation}
and 
\begin{equation}
	\begin{split}
		\label{Eq:F4}
		F^{(4)} &= g_0\alpha_{s,0}(c a^\dag - c^*a), \\
		c &\equiv \beta \cosh(r) + \beta^{*} e^{i2\theta} \sinh(r), \\
		\tilde\Delta_r(\alpha) &\equiv [\omega_r'(\alpha) + \proj{S(\alpha)} - \omega_p]/\cosh(2r), \\
		\delta &= \omega_{1,0}'''-\omega_s.
	\end{split}
\end{equation}
In obtaining this master equation, we have moved to a frame rotating at $\omega_{1,0}'''$ for the qubit and at $\omega_p$ for the resonator. We have also neglected rapidly oscillating terms in $H_{\mathrm{SB}}^{(4)}$ using the rotating wave approximation. 

Finally and as presented in more details in Appendix~\ref{annsec:squeezon_bogoliubov_transformation}, we have assumed that the photon population in the transformed frame is small. As can be seen from the term $\sinh^2(r)\sD[\ad]\rho$ responsible for heating in \eqref{eqn:master_equation_4}, this assumption will only be true in the limit of small squeezing. The squeezing coefficient $r$ is given by the solution of the equations
\begin{subequations}
	\label{eqn:condition_proj_r_proj_theta}
	\begin{align}
		\cos\lsb\arg(\Upsilon_p) - 2\theta\rsb &= \frac{\omega_r'(\alpha) + \proj{S(\alpha)} - \omega_p}{2|\Upsilon_p|} \tanh2r, \\
		\sin\lsb\arg(\Upsilon_p) - 2\theta\rsb &= \frac{\kappa}{4|\Upsilon_p|} \sinh(2r).
	\end{align}
\end{subequations}
While complicated and not reproduced here, the solution of these equations is analytical, and yields a maximum squeezing coefficient
\begin{equation}
	r_\mathrm{max} = \frac12 \arcsinh\inp{\frac{4|\Upsilon_p|}{\kappa}}.
\end{equation}
Maximal squeezing is reached for $\omega_p = \omega_r'(\alpha) + \proj{S(\alpha)}$. For the parameters of interest here and in Ref.~\cite{ong:2013a}, $r_\mathrm{max}\sim 0.5$ for all qubit states corresponding to $\sinh^2 r_\mathrm{max} \sim 0.3$. Although we would have prefered to have $\sinh^2r \ll 1$ to justify our approximation, we will see below that this model nevertheless semi-quantitatively compares with experimental results.

% subsection bogoliubov_transformation (end)
%&&&&&&&&&&&&&&&&&&&&&&&&&&&&&&&&&

%&&&&&&&&&&&&&&&&&&&&&&&&&&&&&&&&&
\subsection{Adiabatic elimination of the resonator} % (fold)
\label{sub:adiabatic_elimination_of_the_resonator}
We now adiabatically eliminate the resonator to obtain a master equation for the qubit only. As described in 
Appendix~\ref{sec:adiabatic_elimination_through_the_projector_formalism}, we use the projector formalism~\cite{Zwanzig1960,Gardiner2000} to obtain the following reduced master equation:
\begin{equation}
	\label{eqn:effective_qubit_master_equation}
	\dot\rho_q = -i\comm{\tilde H}{\rho_q} + \tilde\gamma_\downarrow\sD[\sigma_-]\rho_q + \tilde\gamma_\uparrow\sD[\sigma_+]\rho_q + \frac{\tilde\gamma_\varphi}{2}\sD[\sigma_z]\rho_q.
\end{equation}
In this expression, we have defined the rates
\begin{subequations}
	\begin{align}
		\label{eq:GammaDownTilde}
		\tilde\gamma_\downarrow &= \gamma_\downarrow''' + |g_0\alpha_{s,0}c|^2 
		\left\{ \left[ L(-\delta)+L(\delta)\right] \sinh^2r + L(-\delta) \right\}, \\
		\label{eq:GammaUpTilde}
		\tilde\gamma_\uparrow &= |g_0\alpha_{s,0}c|^2 \left\{ \left[ L(-\delta)+L(\delta)\right] \sinh^2r + L(\delta) \right\}, \\
		\tilde\gamma_\varphi &= \gamma_\varphi''',
	\end{align}
\end{subequations}
and the Hamiltonian
\begin{equation}
	\tilde H = \frac{\tilde\delta}{2} \sigma_z +  g_0 (\alpha_{s,0} \sigma_+ + \alpha_{s,0}^* \sigma_-),
\end{equation}
where $L(\delta) = \RealPart[f(\delta)]$ is a Lorentzian with 
\begin{equation}
	f(\omega) = \frac{\kappa/2 + i[\tilde\Delta_r(\alpha) + \omega]}{{\kappa}^2/4 + [\tilde\Delta_r(\alpha)+\omega]^2}.
\end{equation}
Moreover, $\tilde\delta = \omega_{1,0}'''(\alpha) - \omega_s + \ImaginaryPart[S_\downarrow(\delta)-S_\uparrow(-\delta)]$ where $S_\downarrow(\delta)$ and $S_\uparrow(-\delta)$ are defined in Appendix~\ref{sec:adiabatic_elimination_through_the_projector_formalism}. The above master equation is one of the central results of this paper and the significance of the various terms is discussed below.

\subsection{Steady state qubit population} % (fold)
\label{sub:steady_state_qubit_population}

We are now almost ready to compare the above model to the experiments presented in Ref.~\cite{ong:2013a}. It is however useful to give some more details of the experiment so as to compute the appropriate qubit observable for comparison with theory. In the experiment presented in Ref.~\cite{ong:2013a}, a transmon qubit~\cite{Koch2007} is coupled to a nonlinear transmission resonator. The nonlinearity is provided by a Josephson junction embedded in the central conductor of the resonator. The resonator is first driven with a pump drive of frequency $\omega_p$ and power $P_p$ pushing the resonator field out of its ground state. The drive is chosen to be long enough for the coupled resonator-qubit system to reach its steady state. Under this steady state resonator field, the qubit is ac-Stark shifted as described in \eqref{eqn:omega_i_third}. This frequency shift is then measured in order to reveal information about the internal resonator field. 

For this, a second (spectroscopy) drive at frequency $\omega_s$ close to the qubit frequency is turned on while the pump field is still present. This spectroscopy drive excites the qubit only if $\omega_s$ matches the shifted qubit transition frequency. Both drives are then turned off for a time longer than the resonator decay time, but shorter than the qubit's relaxation time. The qubit state is then measured using the standard bifurcation readout procedure~\cite{Mallet2009}. This process is then repeated multiple times for different spectroscopy frequency $\omega_s$ and pump power $P_p$ to extract the probability $P(\ket{1})$ of the qubit to be in its excited state.

To compare our model to experimental results, we therefore compute from \eqref{eqn:effective_qubit_master_equation} the steady state probability of the qubit to be in its excited state. We find:
\begin{equation}
	\label{eqn:mean_proj1_qubit}
	P(\ket{1}) = \frac{\mean{\proj{1,1}}_\mathrm{eq}\inp{\tilde\gamma_2^2+\tilde\delta^2} + 2\tilde\gamma_2\invb{g_0\alpha_{s,0}}^2/(\tilde\gamma_\uparrow+\tilde\gamma_\downarrow)}{\insb{\inp{\tilde\gamma_2^2+4\tilde\gamma_2\invb{g_0\alpha_{s,0}}^2/(\tilde\gamma_\uparrow+\tilde\gamma_\downarrow) }+\tilde\delta^2}},
\end{equation}
where we have defined
\begin{equation}
	\begin{split}
		\tilde\gamma_2 &\equiv \tilde\gamma_\varphi + \frac{\tilde\gamma_\uparrow+\tilde\gamma_\downarrow}{2}, \\
		\mean{\proj{1,1}}_\mathrm{eq} &\equiv \frac{\tilde\gamma_\uparrow}{\tilde\gamma_\uparrow+\tilde\gamma_\downarrow}.
	\end{split}
\end{equation}

The above expression for $P(\ket{1})$ can be understood by focussing on three different contributions, leading to three peaks in $P(\ket{1})$ versus pump power $P_p$ and spectroscopy frequency $\omega_s$. The first peak is obtained for $\tilde\gamma_\downarrow \gg \tilde\gamma_\uparrow$ such that $\mean{\proj{1,1}}_\mathrm{eq} \sim 0$. In this regime, $P(\ket{1})$ reduces to 
\begin{equation}
	P(\ket{1}) \approx \frac{2\tilde\gamma_2\invb{g_0\alpha_{s,0}}^2/\tilde\gamma_\downarrow}{\insb{\inp{\tilde\gamma_2^2+4\tilde\gamma_2\invb{g_0\alpha_{s,0}}^2/\tilde\gamma_\downarrow }+\tilde\delta^2}}.
\end{equation}
That is, we find a Lorentzian centered at $\tilde\delta=0$ with width $\tilde\gamma_2$ in the absence of power broadening. This Lorentzian is the main qubit line and is power-broadened by the spectroscopy field $\alpha_{s,0}$. 

The two other contributions are found when $\tilde\gamma_\uparrow$ is large. From the definition of $\tilde\gamma_\uparrow$ in \eqref{eq:GammaUpTilde}, this requires that $\pm\tilde\delta \sim \tilde\Delta_r(\alpha)$ which corresponds to $L(\pm\delta)$ taking its maximal value. If $|\delta|$ is sufficiently large and if the undriven decay rate $\gamma_\downarrow'''$ is negligible compared to the effective rates arising from squeezing [$\propto |g_0\alpha_{s,0}c|^2/\kappa$ in Eqs.~\eq{eq:GammaDownTilde} and \eq{eq:GammaUpTilde}], then the dominant contribution is $\mean{\proj{1,1}}_\mathrm{eq}$ 
\begin{equation}
	\label{eqn:P_1_sidebands}
	P(\ket{1}) \approx \mean{\proj{1,1}}_\mathrm{eq} \approx \frac{L(-\delta)\sinh^2r + L(\delta)(\sinh^2r+1)}{[L(-\delta)+L(\delta)](2\sinh^2r+1)}.
\end{equation}
In this situation, corresponding to the resolved sideband limit, $P(\ket{1})$ takes the form of red and blue sidebands on either side of the main qubit line. Interestingly these two sidebands depend on the squeezing parameter $r$. In the next section, we compare this three-peak excitation spectra to experimental data and analyze the amplitude, width and position of the peaks to extract information about the internal resonator state and in particular about the squeezing parameter.

% subsection adiabatic_elimination_of_the_resonator (end)
%&&&&&&&&&&&&&&&&&&&&&&&&&&&&&&&&&
% section squeezing_and_sidebands (end)
%&&&&&&&&&&&&&&&&&&&&&&&&&&&&&&&&&

%&&&&&&&&&&&&&&&&&&&&&&&&&&&&&&&&&
\section{Comparison to experiments} % (fold)
\label{sec:sidebands_results}

As discussed in \paperi and in Ref.~\cite{Ong2011}, only the centre qubit line is observed under low spectroscopy power $P_s$. This can be understood from the expressions \eq{eq:GammaDownTilde} and \eq{eq:GammaUpTilde} for $\tilde\gamma_\uparrow$ and $\tilde\gamma_\downarrow$. Indeed, for negligible spectroscopy power $|\alpha_{s,0}|^2\rightarrow 0$ and $\tilde\gamma_\downarrow \gg \tilde\gamma_\uparrow$ such that as discussed above only the centre line is apparent. The position and width of this peak were analyzed in details in \paperi and in Ref.~\cite{Ong2011}.

\begin{figure}
	\centering
	\includegraphics[width=\hsize]{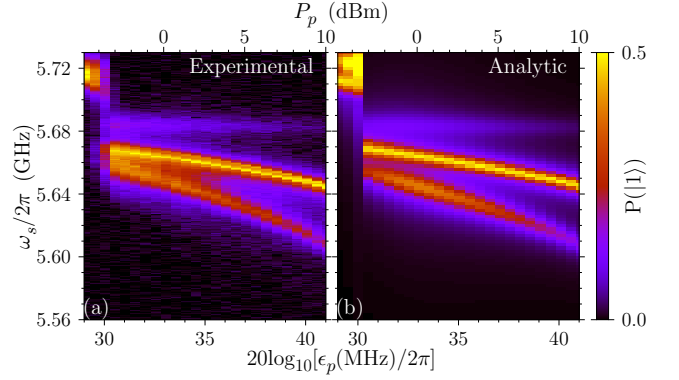}
	\caption{ Qubit spectrum for a strong spectroscopy amplitude $\epsilon_s$ as a function of the pump amplitude $\epsilon_p$, in logarithmic scale. Experimental results (left) are compared to the analytical spectrum (right). Results are for pump frequency $\omega_p/2\pi=6439$~MHz, corresponding to $\Omega/\Omega_C=1.74$ and spectroscopy power $P_s=-4$~dBm, corresponding to $\epsilon_s/2\pi\approx 25$~MHz. All other system parameters can be found in Ref.~\cite{ong:2013a}.}
	\label{fig:sidebands_spectra}
\end{figure}
Here we focus on the situation where the spectroscopy power $P_s$ is important such that $\tilde\gamma_\uparrow$ cannot be neglected with respect to $\tilde\gamma_\downarrow$. The experimental results in this situation for $P(\ket{1})$ versus pump power and spectroscopy frequency -- first presented in Ref.~\cite{ong:2013a} -- are reproduced in Fig.~\ref{fig:sidebands_spectra}a). On the right side of Fig.~\ref{fig:sidebands_spectra}, we show the result of $P(\ket{1})$ as given by the fully analytical expression of \eqref{eqn:mean_proj1_qubit}. In both cases, we see a sudden jump in the qubit transition frequency. This corresponds to switching of the resonator state from its $L$ to its $H$ state and was discussed in details in Ref.~\cite{Ong2011}.  At larger pump power, sidebands are clearly resolved and the  agreement between experiments and theory is excellent. In producing this figure, all parameters except one have been extracted independently and are given in Ref.~\cite{ong:2013a} and \paperi. The adjustable parameter is an ad-hoc multiplicative coefficient to the effective sideband driving amplitude $F^{(4)}$ defined in \eqref{Eq:F4}. This parameter accounts for the large dependence of the coefficient $c$ in $F^{(4)}$ on the squeezing coefficient $\xi=re^{2i\theta}$. For our model to reproduce the experimental sideband amplitudes, we have multiplied $c$ by 2. Considering the number of approximations that are done in obtaining the model, we consider such a factor to be an acceptable correction. It is important to stress that this correction does not affect the position (frequency) of any of the linesmadeMoreover, while it changes the  absolute amplitude of the sidebands relative to the main line, it does not change their width nor does it change the ratio of their amplitude since both sidebands are affected in the same way by this correction.

\begin{figure}
	\centering
	\includegraphics[width=\hsize]{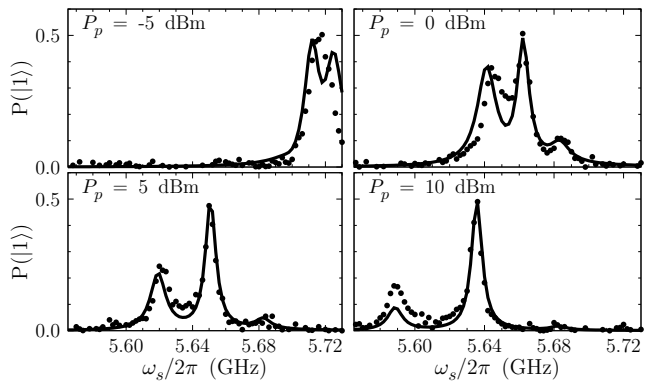}
	\caption{ Line cuts of Fig.~\ref{fig:sidebands_spectra} for four pump powers. Black lines are theoretical predictions. Dots are experimental data.}
	\label{fig:sidebands_spectra_2d}
\end{figure}
For a more quantitative comparison, we show in Fig.~\ref{fig:sidebands_spectra_2d} four linecuts of Fig.~\ref{fig:sidebands_spectra} (a) and (b), for increasing pump power $P_p$. In these plots, the dots correspond to the experimental data, while the lines are the analytical predictions. With the above single correction, the agreement is almost quantitative at all powers. We note however that at $P_p = -5$~dBm (top left panel) the analytical results predict two peaks while a single one is observed experimentally~\footnote{This sideband was observed experimentally with another sample, see the Supplemental Material for Ref.~\cite{ong:2013a}}. In fact, in all cases the sidebands are better resolved in the analytical model than is observed experimentally, especially close to the bifurcation threshold. We attribute this discrepancy to the rotating wave approximation made in Appendix~\ref{sec:adiabatic_elimination_through_the_projector_formalism} where we neglected terms oscillating at the sideband-detuning frequency. As is discussed there and observed in Fig.~\ref{fig:sidebands_spectra_2d}, making this approximation corresponds to considering the resolved-sideband limit. 

\begin{figure}
	\centering
	\includegraphics[width=\hsize]{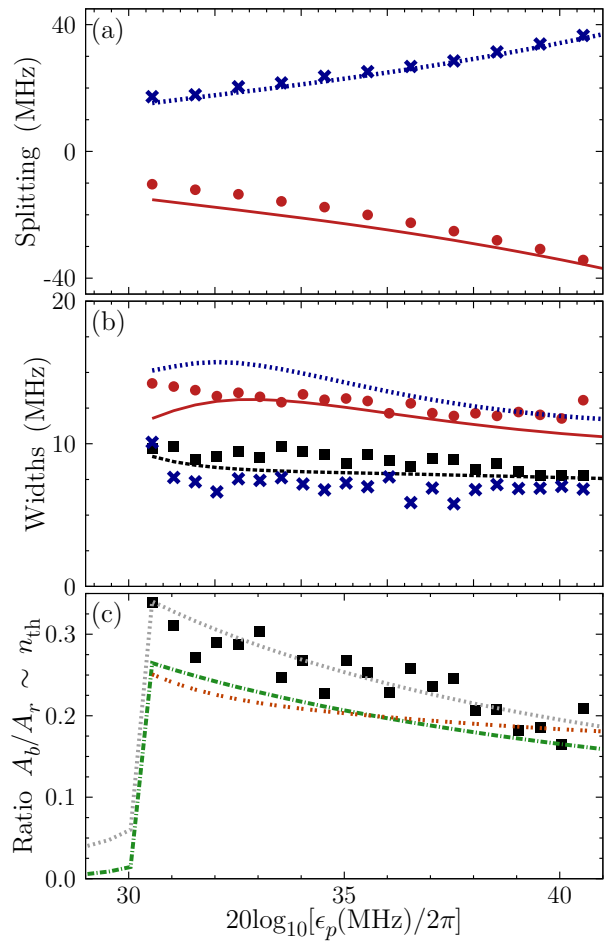}
	\caption{ Sideband splitting $f_i-f_c$ (a), width $w_i$ (b) and ratio of amplitude $A_b/A_r$ (c) as a function of the pump amplitude $\epsilon_p$, in logarithmic scale. In (a) and (b), red circles (full red lines), blue crosses (dotted blue lines) and black squares (dashed black lines) are experimental data (theoretical predictions) for the red sideband ($i=r$), blue sideband ($i=b$) and main line ($i=c$), respectively.  In (c), black squares (orange double-dotted line) are extracted from fits to the experimental (analytical) spectrum, while the dotted grey line and dashed-dotted green line correspond to \eqref{eqn:ratio_1} and \eqref{eqn:ratio_2}, respectively.}
	\label{fig:sidebands_analysis}
\end{figure}
While the analytical expression for $P(\ket{1})$ in terms of $\omega_s$ is not simply that of three superposed Lorentzians, it is useful to compare fits of both the analytical expressions and the experimental data to such a simplified model. The position, amplitude and widths of the three peaks were therefore extracted by fitting the sum of three Lorentzian curves both to the experimental and analytical spectra. We plot the results of these fits in Fig.~\ref{fig:sidebands_analysis}, where we show the sideband detuning from the main line [panel (a)], the width of the three peaks [panel (b)], and the ratio of amplitude of the blue to the red sidebands [panel (c)]. Due to the breakdown of the resolved-sideband approximation and as mentioned above, we see in (a) that the sidebands are slightly more separated in the analytical model (lines) than in the experimental data (dots). We see from panel (b) that the width of the red sideband (full line and circles) and the centre peak (dashed line and squares) are in quantitative agreement. The model however shows a blue sideband that is much wider than obtained experimentally. We attribute this discrepancy to the small signal-to-noise ratio of the blue sideband (i.e.~it is not very distinguishable from the noise). In fact, the statistical error on the fit parameters of the blue sidebands are about 30\%. 

In panel (c), we show the ratio of the amplitude of the blue sideband to that of the red sideband. In addition to the data from the fit to the experimental data (black squares) and to the analytical expression (orange double-dotted line), we have extracted this ratio directly from the analytical expression for $P(\ket{1})$, i.e. without assuming a Lorentzian profile. Since the blue (red) sideband is at a frequency corresponding to $\delta=\tilde\Delta_r$ ($\delta = -\tilde\Delta_r$), the ratio of amplitudes is given by $A_b/A_r = P(\ket{1})|_{\delta = -\tilde\Delta_r}/P(\ket{1})|_{\delta = \tilde\Delta_r}$. Using the simplified expression \eqref{eqn:P_1_sidebands} for the amplitude of the sidebands, we find
\begin{subequations}
	\begin{align}
		\label{eqn:ratio_1}
		\frac{A_b}{A_r}
		%\frac{P(\ket{1})|_{\delta = -\tilde\Delta_r}}{P(\ket{1})|_{\delta = \tilde\Delta_r}} 
		&\approx\frac{[L(\tilde\Delta_r)+L(-\tilde\Delta_r)]\sinh^2 r + L(-\tilde\Delta_r) }{[L(-\tilde\Delta_r)+L(\tilde\Delta_r)]\sinh^2 r + L(\tilde\Delta_r)} \\
		\label{eqn:ratio_2}
		&\approx \frac{\sinh^2r}{\sinh^2r + 1},
	\end{align}
\end{subequations}
where in the last approximation, we have assumed that $L(-\tilde\Delta_r) \gg L(\tilde\Delta_r)$, which is valid in the limit $\tilde\Delta_r \gg \kappa$, i.e. in the resolved-sideband limit. \eqref{eqn:ratio_1} above corresponds to the dotted grey line in panel (c) and \eqref{eqn:ratio_2} to the dashed-dotted green line. Because of space constrains, only the simpler expression \eqref{eqn:ratio_2} was used for comparison to the experimental data in Ref.~\cite{ong:2013a}.

We emphasize once more that the correction factor applied to the theoretical results does not affect any of the results showed in Fig.~\ref{fig:sidebands_analysis}. As a result, given the quantitative agreement displayed in panel c, it is possible to accurately determine the squeezing coefficient $r$ of the \emph{intra}-resonator field. As already mentioned above, we find a maximal value of $\sinh^2 r_\mathrm{max} \sim 0.3$ corresponding to $r_\mathrm{max}\sim 0.5$. 

Alternatively, the ratio of the two sidebands also allows us to extract the effective temperature of the oscillator as described in the quadruply transformed frame by the master equation \eq{eqn:master_equation_4}. Indeed, the last line of this master equation takes the standard form
\begin{equation}
\kappa (n_\mathrm{th}+1)\sD[a]\rho^{(4)} + \kappa n_\mathrm{th}\sD[\ad]\rho^{(4)},
\end{equation}
where we have identified $\sinh^2(r)$ with an effective thermal occupation number of the oscillator, $n_\mathrm{th} = \sinh^2(r)$. Using the Bose-Einstein distribution for $n_\mathrm{th}$ then allows us to define an effective temperature $T_\mathrm{eff}$ for the system due to squeezing of the resonator field. This corresponds to the essential prediction of the quantum heating theory~\cite{Drummond1980,Marthaler2006,dykman:2011a,andre:2012a}. As discussed in more details in Ref.~\cite{ong:2013a}, $T_\mathrm{eff}$ corresponds to a few tenths of quanta, much larger than the temperature expected from the base temperature of the dilution refrigerator and filtering of the lines. Moreover, the non-monotonic dependence with pump power, especially the decrease of the effective temperature with increasing pump power above bifurcation exclude a classical heating effect (due to the contacts for example) from being the cause of this effective temperature.

% section sidebands_results (end)
%&&&&&&&&&&&&&&&&&&&&&&&&&&&&&&&&&

%&&&&&&&&&&&&&&&&&&&&&&&&&&&&&&&&&
\section{Conclusion} % (fold)
\label{sec:conclusion}
We have developed a theoretical model for a multi-level qubit coupled to a pumped nonlinear resonator. The model holds within the dispersive regime of circuit QED and for pumping powers well above the bifurcation threshold of the resonator. The reduced qubit master equation that we have obtained contains Purcell decay, measurement-induced dephasing, dressed dephasing, quadratic ac-Stark shift, Lamb shift, as well as the first order correction of the resonator's squeezing on the qubit. This allows us to obtain quantitative agreement with experimental data in a wide range of parameters, without adjustable parameters. In this way, we show how the qubit can be used as a probe of  squeezing of the intra-resonator field. By comparing the ratio of the red and blue sidebands in the qubit excitation spectrum, we have extracted the squeezing coefficient of the field or equivalently the effective temperature of the so-called quasi-oscillator, providing a direct demonstration of quantum heating. 

Interesting extensions to this model include a finer treatment of the intra-resonator squeezing, especially the squeezing angle, as well as a qubit-dependent squeezing. This latter aspect could provide insights on how to improve dispersive qubit measurement using squeezing, or on how measurement-induced dephasing is changed by squeezing of the intra-resonator field. 
% section conclusion (end)
%&&&&&&&&&&&&&&&&&&&&&&&&&&&&&&&&&

\emph{Note added.--} After this work was completed, we became aware of related work~\cite{peano:2013a}.

%&&&&&&&&&&&&&&&&&&&&&&&&&&&&&&&&&
\begin{acknowledgments}
We acknowledge discussions with J.~Gambetta, M.~Dykman and within the Quantronics group. We acknowledge support from NSERC, FQRNT, the Alfred P. Sloan Foundation, CIFAR, the European project Scaleqit and the ITN network CCQED and the Australian Research Council Centre of Excellence Scheme (Grant No. EQuS CE110001013). We thank Calcul Qu\'ebec and Compute Canada for computational resources.
\end{acknowledgments}
%&&&&&&&&&&&&&&&&&&&&&&&&&&&&&&&&&

\appendix
%&&&&&&&&&&&&&&&&&&&&&&&&&&&&&&&&&
\section{Squeezon transformation} % (fold)
\label{annsec:squeezon_bogoliubov_transformation}
In this Appendix, we diagonalise the sideband Hamiltonian by applying the transformation $\tS$ of \eqref{eqn:tS} on the master equation \pref{eqn:master_equation_third_sb}. Given the action of $\tS$ on $a$ [see \eqref{eqn:a_S}], we find for the transformed dissipator
\begin{equation}
	\label{eqn:squeezed_photon_loss}
	\begin{split}
		\sD[a]\rho &\rightarrow \cosh r\sD[a]\rho + \sinh r\sD[\ad]\rho \\
		&\quad -i\frac{\sinh(2r)}{4}\comm{-i(e^{-i2\theta}a^2 - e^{2i\theta}{\ad}^2)}{\rho} \\
		&\quad -\cosh r \sinh r\lcb e^{-i2\theta} a\comm{\rho}{a} + \comm{\ad}{\rho}\ad e^{2i\theta}\rcb.
	\end{split}
\end{equation}
As we can see in the first line, what was pure damping in the original frame now sees heating in the transformed frame. The second line of the above expression takes the form of a commutator and can be added to the Hamiltonian part of the master equation. Finally, we will neglect the last line because it vanishes if the resonator is in its ground state. Because of the presence of the heating term, this is an approximation that restricts this theory to low squeezing (i.e.~low effective temperature). 

Taking into account the above contribution from the dissipation, we can then transform the system Hamiltonian $H_s''' = H_0'''+H_2'''+H_\mathrm{SB}'''$, yielding
\begin{widetext}
	\begin{equation}
		\begin{split}
			H_s^{(4)} &= H_0''' + \lp {F^{(4)}}^\dag \sigma_- + F^{(4)} \sigma_+\rp \\
			&\quad + \omega_r'(\alpha)\ada + \proj{S(\alpha)}\ada + \insb{2\sinh^2r\inp{\omega_r'(\alpha) + \proj{S(\alpha)}-\omega_p} - \sinh(2r)\inp{\Upsilon_p e^{-i2\theta} + \hc}}(\ada+\tfrac12) \\
			&\quad - \insb{\frac{\sinh(2r)}{2}\inp{\omega_r'(\alpha) + \proj{S(\alpha)}-\omega_p} - \Upsilon_p e^{-2i\theta} \sinh^2r - \Upsilon_p^* \cosh^2re^{i2\theta} + i \frac{\kappa\sinh(2r)}{4}  } e^{-i2\theta} a^2 +\hc,
		\end{split}
	\end{equation}
\end{widetext}
where $F^{(4)}$ is 
\begin{equation}
	F^{(4)} = g_0\alpha_{s,0}\lsb\lp \beta \cosh r + \beta^* e^{i2\theta}\sinh r \rp\ad - \hc\rsb.	
\end{equation}
We see that choosing $r$ and $\theta$ such that
\begin{equation}
	\label{eqn:condition_r_theta}
	\begin{split}
		-(\omega_r'(\alpha) + \proj{S(\alpha)} - \omega_p) \frac{\sinh(2r)}{2} + i\frac{\kappa\sinh(2r)}{4} \\
		+ \Upsilon_p e^{-2i\theta}\cosh^2r + \Upsilon_p^* e^{2i\theta}\sinh^2r  = 0,
	\end{split}
\end{equation}
yields vanishing squeezing terms, leaving only a renormalized harmonic oscillator and a driven qubit. The solution to this equation can be expressed in the simpler form of \eqref{eqn:condition_proj_r_proj_theta} when considering the real and imaginary parts separately. 

Assuming that the squeezing coefficient does not depend on the qubit state, and moving to a frame rotating at $\omega_p$ for the resonator and $\omega_{1,0}'''$ for the qubit, we then obtain the master equation~\pref{eqn:master_equation_4}.

% annsection squeezon_bogoliubov_transformation (end)
%&&&&&&&&&&&&&&&&&&&&&&&&&&&&&&&&&

%&&&&&&&&&&&&&&&&&&&&&&&&&&&&&&&&&
\section{Adiabatic elimination through the projector formalism} % (fold)
\label{sec:adiabatic_elimination_through_the_projector_formalism}
In this Appendix we use the projector formalism~\cite{Zwanzig1960,Gardiner2000} to adiabatically eliminate the resonator's degrees of freedom. We first note that the master equation \eq{eqn:master_equation_4} can be expressed as
\begin{equation}
	\dot\rho = \sL_r\rho + \sL_c(t)\rho + \sL_q\rho,
\end{equation}
where to simplify the notation used in this Appendix we have dropped the index on $\rho$ and defined the resonator, qubit and coupling Lindbladians as
\begin{equation}
	\begin{split}
		\sL_q\rho_q &\equiv -i\comm{H_0^{(4)}}{\rho_q} + \gamma_\downarrow'''\sD[\sigma_-]\rho_q + \frac{\gamma_\varphi'''}{2} \sD[\sz]\rho_q, \\
		\sL_r\rho_r &\equiv -i\comm{H_2^{(4)}}{\rho_r} + \kappa\insb{1+\sinh^2(r)}\sD[a]\rho_r \\
		&\quad + \kappa\sinh^2(r)\sD[\ad]\rho_r, \\
		\sL_c(t)\rho &\equiv -i\comm{H_{\mathrm{SB}}^{(4)}}{\rho},
	\end{split}
\end{equation}
respectively. We then assume that the resonator relaxes much faster than the qubit (i.e. $\kappa\gg\gamma,\gamma_\varphi$). This allows us to assume a stationary state for the resonator given by $\dot\rho_r^s = \sL_r \rho_r^s = 0$. In the same way, we also assume that the qubit-resonator state can be written as
\begin{equation}
	\rho(t) = \rho_r^s\otimes\rho_q(t),
\end{equation}
where $\rho_q(t)$ is the reduced qubit density matrix. In order to find the evolution equation for $\rho_q(t)$, we define the projector on the qubit subspace $\sQ$:
\begin{equation}
	\sQ\rho \equiv \rho_r^s\otimes\trace[r]{\rho},
\end{equation}
and the complementary projector $\sR \equiv \sI - \sQ$, where $\sI$ is the identity. There are a number of useful identities that one can prove with these projectors. Among these are $\sL_r\sQ = \sQ\sL_r = 0$, $\sL_q\sQ = \sQ\sL_q$, $\sR\sL_r = \sL_r\sR$ and $\sR\sL_q = \sL_q\sR$. The first property arises from the adiabatic approximation and the trace-preserving nature of Lindbladians, while the three others are consequences of the definition of $\sQ$ and $\sR$~\cite{Gardiner2000}. 

Using the above properties, and assuming --- without loss of generality --- that $\sQ\sL_c(t)\sQ = 0$, we then search for the evolution equations of $v(t) = \sQ\rho(t)$ and $w(t) = \sR\rho(t)$. We obtain
\begin{equation}
	\begin{split}
		\dot v(t) &= \sQ\sL_c(t) w(t) + \sL_q v(t), \\
		\dot w(t) &= (\sL_r + \sR\sL_c(t) + \sL_q) w(t) + \sR\sL_c(t) v(t).
	\end{split}
\end{equation}
Assuming that $\sL_r$ contains the dominant dynamic and that the integration time is long compared to $1/\kappa$ yet short compared to $1/\gamma,1/\gamma_\varphi$, we obtain the approximate solution for $w(t)$
\begin{equation}
	w(t) \approx \int_0^\inf d\bar t \exp\insb{\sL_r \bar t} \sR\sL_c(t-\bar t) v(t).
\end{equation}
We can then replace this solution into the equation for $v(t)$ and, using the definition of $\sQ$, obtain an equation for $\dot\rho_q$
\begin{equation}
	\dot\rho_q = \sL_q \rho_q(t) + \int_0^\inf dt' \trace[r]{\sL_c(t) e^{\sL_r t'} \sL_c(t-t') \rho(t)}.
\end{equation}
Given the form of $\sL_c(t)$, one can then show that the qubit reduced master equation can be written as
\begin{equation}
	\begin{split}
		\dot\rho_q &= \sL_q\rho_q - i \comm{\delta H}{\rho_q} \\
		&\quad + \RealPart[S_\uparrow(-\delta)] \sD[\sigma_+]\rho_q + \RealPart[S_\downarrow(\delta)]\sD[\sigma_-]\rho_q,		
	\end{split}
\end{equation}
where we have defined the spectra
\begin{equation}
	\begin{split}
		S_\uparrow(\omega) &= \int_0^\inf dt' e^{i\omega t'} \mean{{F^{(4)}}^\dag(t') F^{(4)}(0)}_s, \\
		S_\downarrow(\omega) &= \int_0^\inf dt' e^{i\omega t'} \mean{{F^{(4)}}(t') {F^{(4)}}^\dag(0)}_s, 
	\end{split}
\end{equation}
with $\mean{\cdot}_s = \trace[r]{\cdot \rho_r^s}$ and where
\begin{equation}
	\delta H = \ImaginaryPart[S_\uparrow(-\delta)]\sigma_-\sigma_+ + \ImaginaryPart[S_\downarrow(\delta)]\sigma_+\sigma_-.
\end{equation}
Importantly, in obtaining the above master equation, we have assumed that we could neglect terms oscillating at a frequency $2\delta$. As discussed in section~\ref{sec:sidebands_results}, this approximation corresponds to the well resolved sideband limit. 

Finally, we use the quantum regression theorem~\cite{Gardiner2000} to obtain
\begin{equation}
	S_\uparrow(\omega) = |g_0\alpha_{s,0}c|^2 \insb{f(\omega) \sinh^2r + f^*(-\omega) (1+\sinh^2r)},
\end{equation}
where we have defined the complex function
\begin{equation}
	\label{eqn:complex_f}
	f(\omega) = \frac{\kappa/2 + i[\tilde\Delta_r(\alpha) + \omega]}{{\kappa}^2/4 + [\tilde\Delta_r(\alpha)+\omega]^2},
\end{equation}
and $S_\downarrow(\omega) = S_\uparrow(\omega)$. With this spectrum, we easily get the reduced master equation~\pref{eqn:effective_qubit_master_equation} for the qubit only.
% section adiabatic_elimination_through_the_projector_formalism (end)
%&&&&&&&&&&&&&&&&&&&&&&&&&&&&&&&&&

%\bibliography{/Users/mboisson/Documents/Articles/Biblio}
\bibliography{Biblio}

\end{document}